\documentstyle[aps,pra,epsfig]{revtex}

\begin{document}

\draft

\title{Superfluid Dynamics \\ 
of a Bose-Einstein Condensate in a Periodic Potential}

\author{C. Menotti$^{1,2}$, A. Smerzi$^{1,3}$ and A. Trombettoni$^{4}$}
\address{$^1$ 
Istituto Nazionale di Fisica per la Materia BEC-CRS and
Dipartimento di Fisica, Universit\`a di Trento, I-38050 Povo, Italy\\
$^2$ Dipartimento di Matematica e Fisica,
Universit\`{a} Cattolica del Sacro Cuore,
I-25121 Brescia, Italy \\
$^3$ Theoretical Division, Los Alamos
National Laboratory, Los Alamos, NM 87545, USA\\
$^4$ Istituto Nazionale per la Fisica della Materia and
Dipartimento di Fisica, Universit\`a di Parma,
parco Area delle Scienze 7A, I-43100 Parma, Italy}

\date{\today}
\maketitle

\begin{abstract}
We investigate the superfluid properties of a Bose-Einstein condensate
(BEC) trapped in a one dimensional periodic potential. We study, both
analytically (in the tight binding limit) and numerically, the Bloch
chemical potential, the Bloch energy and the Bogoliubov dispersion
relation, and we introduce {\it two} different, density dependent, effective
masses and group velocities. 
The Bogoliubov spectrum predicts
the existence of sound waves, and the arising of energetic and
dynamical instabilities at critical values of the BEC quasi-momentum
which dramatically affect its coherence properties.
We investigate the dependence of the dipole and Bloch
oscillation frequencies in terms of an effective mass averaged
over the density of the condensate. 
We illustrate our results with several animations obtained solving
numerically the time-dependent Gross-Pitaevskii equation.
\end{abstract}
%\pacs{PACS: 63.20.Pw, 05.45.-a}

%\begin{multicols}{2}

\section{INTRODUCTION} 

The study of the superfluid properties of Bose-Einstein condensates
(BECs) trapped in periodic potentials are attracting a fast growing
interest. The main reason is that the control parameters of such
systems are widely tunable in realistic experiments, allowing for the
investigation of different and fundamental issues of quantum
mechanics, ranging from quantum phase transitions \cite{greiner02} and
atom optics \cite{rolston02,orzel01} to the dynamics of Bloch and
Josephson oscillations
\cite{anderson98,morsch01,cataliotti01}. Several efforts are also
focusing on the realization of new technological devices as BEC
interferometers working at the Heisenberg limit \cite{orzel01}, and
quantum information processors \cite{rolston02}.

The dynamics of BECs in lattices is highly non-trivial, essentially
because of the competition/interplay between the {\em discrete}
translational invariance of the periodic potential and the {\em
nonlinearity} arising from the interatomic interactions.  For deep
enough optical potentials, interactions induce a quantum transition
from the superfluid to a Mott-insulator phase
\cite{greiner02,fisher,jaksch}. In this work we will study the system
in a region of parameters such that its ground state stands deeply in
the superfluid phase, with the dynamics governed by the
Gross-Pitaevskii equation (GPE).  Because of the discrete
translational invariance, the excitation spectrum of the system
exhibits a band structure which has several analogies with the
electron Bloch bands in metals
\cite{berg98,javanainen99,chiofalo00}. On the other hand, the
coexistence of Bloch bands and nonlinearity allows, for instance,
solitonic structures \cite{trombettoni01,abdullaev01,rey03} and
dynamical instabilities \cite{wu01,smerzi02,machholm03} which do not
have an analog neither in metals, nor in Galilean invariant systems.
   
Exact, time-dependent solutions of the GPE with an external periodic
potential, Eq.(\ref{GPE}), can be written as Bloch states, namely as
plane waves modulated by functions having the same periodicity of the
lattice.  The dynamics of small amplitude perturbations on top of
these states satisfies two coupled, linear Bogoliubov equations, which
can be solved numerically. However, when the interwell barriers of the
periodic potential are high enough, the system can be described in a
nonlinear tight binding approximation and several important properties
of the system can be retrieved analytically \cite{smerzi03-gen}.
Indeed, in the nonlinear tight binding approximation the continuous
Gross-Pitaesvkii equation can be replaced with a
discrete nonlinear equation (DNL), Eq.(\ref{DNLS_gen}), where the relevant
observables of the system are the number of particles $N_l(t)$ trapped
in well $l$ and the relative phases
$\phi_{l,l+1}(t)=\phi_{l+1}(t)-\phi_{l}(t)$.  In this paper, we
rewrite the results derived in \cite{smerzi03-gen} in a more
convenient form, namely in terms of {\it two} effective masses and
group velocities.  Furthermore, we compare our analytical expressions
with full numerical solutions, and we extend our analysis to
investigate the behaviour of the system at low optical potential
depths, where the nonlinear tight binding approximation breaks down.
We show that the phenomena predicted by the DNL equation 
(\ref{DNLS_gen}) can be generalized to the case of shallow
potentials, bringing new insights on the dynamics of the system.

\section{DISCRETE NONLINEAR DYNAMICS} 

In the ``classical" (mean field) approximation, the BEC dynamics at $T=0$ 
is governed by the Gross-Pitaevskii equation \cite{dalfovo99}

\begin{equation}
\label{GPE}
i \hbar \frac{\partial \Psi}{\partial t}({\vec r},t)= 
\left[-\frac{\hbar^2 \nabla^2 }{2 m}  + V_{ext}({\vec r}\,) + 
g \mid \Psi({\vec r},t) \mid^2 \right] \Psi({\vec r},t)=
\mu  \Psi({\vec r},t),
\end{equation}
where $g=4 \pi \hbar^2 a/m$, with $m$ the atomic mass and $a$ the
$s$-wave scattering length: $a>0$ ($a<0$) corresponds to an effective
interatomic repulsion (attraction). In the following we consider only
a BEC with repulsive interatomic interactions. The external potential
$V_{ext}$ includes the optical periodic potential $V_P$, which is
typically superimposed to an harmonic (or linear) potential $V_{M}$.
The periodic potential is

\begin{equation}
\label{laser_potential}
V_P=s E_R \sin^2 \left(\frac{\pi x}{d}\right),
\end{equation}
where $d$ is the lattice spacing and $\pi/d$ is the wavevector of the
lasers in the lattice direction.  The lattice spacing determines the
Bragg momentum

\begin{equation}
q_B \, = \, \hbar \, \frac{\pi}{d},
\label{qB}
\end{equation}
corresponding to the boundary of the first Brillouin zone.  The energy
barrier between adjacent sites is expressed in units of the recoil
energy $E_R=q_B^2/ 2m$. From (\ref{laser_potential}) we see that the
minima of the laser potential are located at the positions $x_l=l d$
($l$ is an integer).  Around these points, $V_P \approx
m\tilde{\omega}_x^2 (x-x_l)^2/2$, where $\hbar \tilde{\omega}_x= 2
\sqrt{s} E_R$.

The harmonic potential is $V_M= m [\omega_x^2 x^2 + \omega_y^2 y^2 +
\omega_z^2 z^2]/2$. Since, typically, $\omega_x \ll \tilde{\omega}_x$,
it is convenient to write the external potential as $V_{ext} = V_L +
V_D$, where the lattice potential $V_L = s E_R \sin^2{(\pi x/d)} + m
[\omega_y^2 y^2 + \omega_z^2 z^2]/2$ includes the transverse confining
field, and the ``driving" field $V_D = m \omega_x^2~ x^2/2$ gives the
effective force acting on the center of mass of the condensate wave
packet.

In order to understand the basic physics of the system, we first
consider the case of deep optical lattices, where analytic solutions
can be obtained in the tight binding approximation. Then we study the
behaviour of the system beyond the tight binding
limit, solving numerically the Gross-Pitaevskii and Bogoliubov
equations with arbitrarily shallow periodic potentials.

As it has been previously shown in \cite{smerzi03-gen}, when the
interwell barriers are much higher than the chemical potential, it is
possible to write the condensate wavefunction as

\begin{equation}
\label{TB_gen}
\Psi(\vec{r},t)= \sum_{l} \psi_l(t) ~ \Phi_l(\vec{r}; N_l(t)),
\end{equation}
where the Wannier wavefunctions $\Phi_l$ are well localized in each
well.  The total number of atoms is $N_T=\sum_l N_l \equiv \sum_l
|\psi_l|^2$.  Replacing this Ansatz in the Gross-Pitaevskii equation
(\ref{GPE}) and integrating out the spatial degrees of freedom, we
find the DNL equation 

\begin{eqnarray}
\label{DNLS_gen}
&& i \hbar \frac{\partial \psi_l}{\partial t} =
{\cal V}_l ~ \psi_l + \mu_l^{loc}~ \psi_l  
- \chi ~ [\psi_l(\psi^\ast_{l+1}+\psi^\ast_{l-1})+c.c.]~ \psi_l + \cr 
&& - ~ [K+\chi~ (\mid \psi_l \mid^2 + \mid \psi_{l+1} \mid^2)]~ \psi_{l+1} 
   - ~ [K+\chi~ (\mid \psi_l \mid^2 + \mid \psi_{l-1} \mid^2)]~ \psi_{l-1}.
\end{eqnarray}
The ``local" chemical potential $\mu_l^{loc}$ is the sum of
three contributions

\begin{equation}
\mu_l^{loc} = \mu_l^{kin} + \mu_l^{pot} + \mu_l^{int}=
\int d\vec{r} ~ \bigg[ \frac{\hbar^2}{2m}
~(\vec{\nabla} \Phi_l)^2 
+ V_L ~\Phi_{l}^2 
+ g |\psi_l|^2 ~\Phi_l^4 \bigg],
\label{mu}
\end{equation}
which depend on the atom number $N_l$ explicitly through $|\psi_l|^2$
and implicitly through the shape of the $\Phi_l$'s.  The tunneling
rates $K_{l,l \pm 1}$ between the adjacent sites $l$ and $l \pm 1$ are

\begin{equation}
K \simeq - \int d\vec{r} ~ \bigg[ \frac{\hbar^2}{2m}
\vec{\nabla} \bar{\Phi}_l \cdot \vec{\nabla} \bar{\Phi}_{l \pm 1}
+ \bar{\Phi}_l V_{ext} \bar{\Phi}_{l \pm 1} \bigg],
\label{kappa}
\end{equation}
where the on-site wavefunctions have been calculated with an average
number of atoms per site, $N_0= \mid \psi_0 \mid^2$, namely
$\Phi_l(N_l) \simeq \bar{\Phi}_l(N_0)$ (a discussion of the validity
of this approximation is in \cite{smerzi03-gen}).  On the same line,
the coefficient $\chi$ is given by

\begin{equation}
\chi = - g \int d\vec{r} ~ {\bar{\Phi}_{l}}^3 \bar{\Phi}_{l \pm 1},
\label{chi}
\end{equation}
and the on-site energies ${\cal V}_l$, arising from any 
external potential superimposed to the optical lattice, are
\begin{equation}
{\cal V}_l = \int d\vec{r}~  V_D~ {\bar{\Phi}}_l^2,
\label{epsilon}
\end{equation}
such that ${\cal V}_l \propto l^2$ (${\cal V}_l \propto l$) when the
driving field is harmonic (linear). 

The dependence of the local chemical potential on the number of atoms
is affected by the effective dimensionality of the condensates trapped
in each well of the lattice. This can be determined comparing the
interaction chemical potential $\mu^{int}_l = |\psi_l|^2 g \int
d\vec{r}~ \Phi_l^4$ and the three frequencies, $\tilde{\omega}_x$,
$\omega_y$, $\omega_z$ obtained expanding the lattice potential around
the minima of each well $V_L \simeq m [\tilde{\omega}_x^2 (x-x_l)^2 +
\omega_y^2 y^2 + \omega_z^2 z^2]/2$.  For instance, when
$\tilde{\omega}_x, \omega_y, \omega_z \gg \mu^{int}_l$, the spatial
width of each trapped condensate does not depend (in first
approximation) on the number of particles $N_l$ in the same well, and
the condensate wavefunction in each valley of the periodic potential
is well approximated by a gaussian. We consider this as a $0D$
(zero-dimensional) case: then, the nonlinear tight binding
approximation (\ref{TB_gen}) reduces to the usual tight binding
approximation $\Psi(\vec{r},t)= \sum_{l} \psi_l(t) ~
\Phi_l(\vec{r}\,)$ \cite{trombettoni01}. 
The $1D$ case arises when two frequencies are
larger than the interaction chemical potential. In this case the
system realizes an array of weakly coupled cigar-shaped condensates,
with $\Phi_l$ factorized as gaussians along the two tight directions
and a Thomas-Fermi in the other direction. In the $2D$ case only one
frequency is smaller than the local interaction chemical potential: we
have an array of pancake-like condensates, with $\Phi_l$ factorized as
a gaussian along the tight direction and a Thomas-Fermi in the two
other directions. The $3D$ case is given by the condition $\mu^{int}_l
\gg \tilde{\omega}_x, \omega_y, \omega_z$ and the wavefunction in the
$l$th well $\Phi_l$ is simply given by a three-dimensional
Thomas-Fermi function.  The crucial point is that the effective
dimensionality of the condensates gives a different scaling of the
local interaction chemical potential (\ref{mu}) with the number of
atoms

\begin{equation}
\mu^{loc}_l =  U_{\alpha} \mid \psi_l \mid^{\alpha},
\label{mu_dimensional}
\end{equation}
with $\alpha = {4 \over{2 + D}}$, where $D=0,1,2,3$, $|\psi_l|^2$ is
the number of atoms in well $l$, and $U_{\alpha}$ is a constant which
does not depend on the number of atoms nor on the site index. When
$\chi N_0 \ll K$ and $D=0$, the DNL equation 
(\ref{DNLS_gen}) gives the discrete nonlinear Schr\"odinger equation
\cite{trombettoni01}.

\section{EXCITATION SPECTRA} 

In this section we derive the Bloch and the Bogoliubov excitation
spectra of the system in absence of any driving field (${\cal V}_l =
0$).  First we derive our results analytically in the tight binding
approximation; then we solve the equations numerically for a wide set
of parameters to extend our treatment beyond the tight binding regime.

\subsection{Bloch energy, Bloch  
chemical potential, effective masses and group velocities}

The Bloch states $\Psi_p(x)=e^{ipx/\hbar} {\tilde \Psi_p}(x)$, where
${\tilde \Psi_p}(x)$ is periodic with period $d$, are exact stationary
solutions of the Gross-Pitaevskii equation (\ref{GPE}).  Both the
energy per particle $\varepsilon_{\alpha}(p)$ (Bloch energy) and the
chemical potential $\mu_{\alpha}(p)$ of such solutions form a band
structure, so that they can be labeled by the quasi-momentum $p$ and
the band index $\alpha$.

The DNL equation (\ref{DNLS_gen}) describes only the lowest band of the
spectrum (in the following, we will consider only the lowest band
$\alpha=1$, and we will omit, for simplicity, the band index $\alpha$). 
Exact
solutions of the DNL equation are the "plane waves" $\psi_l= \psi_0~e^{i(p l d-
\mu t)/\hbar}$, where $p$ is the quasi-momentum, and $l$ is the site
index (note that the $\psi_l$ are plane waves in the discrete
$l$-space, but do not correspond to plane waves in real space).
Within the DNL equation framework, the energy per
particle $\varepsilon(p)$ and chemical potential $\mu(p)$
corresponding to these solutions are

\begin{eqnarray}
\varepsilon (p)&=& \varepsilon^{loc} - 
2~(K + 2~ \chi~N_0)
\cos \left({ \pi p \over q_B}\right) \equiv
\varepsilon^{loc} - {q_B^2 \over \pi^2 m_\varepsilon} 
\cos \left({ \pi p \over q_B}\right), \\
\mu (p)&=& \mu^{loc} - 2~(K + 4~ \chi~N_0) 
\cos \left({ \pi p \over q_B}\right) \equiv
\mu^{loc} - {q_B^2 \over \pi^2 m_\mu} 
\cos \left({ \pi p \over q_B}\right) ,
\label{chempot}
\end{eqnarray}
where $\mu^{loc} = \mu^{loc}_l|_{\psi_l=\psi_0}= \partial (N_0
\varepsilon^{loc}) / \partial N_0$, with $N_0 = |\psi_0|^2$ the number
of atoms per well and $\varepsilon^{loc}=2 U_\alpha
N_0^{\alpha/2}/(\alpha+2)$.  In the previous equations we have
introduced the effective masses $m_\varepsilon$ and $m_\mu$, to
emphasize the low momenta (long wavelength) quadratic behaviour of the
Bloch energy spectrum and of the chemical potential
\cite{menotti1}. It turns out that several dynamical properties of the
system can be intuitively understood in terms of such effective
masses. This approach is quite common, for instance, in the theory of
metals, where $m_\mu \equiv m_\varepsilon$.  However in BEC, because
of the nonlinearity of the Gross-Pitaevskii equation, the two relevant
energies of the system, $\varepsilon$ and $\mu$, have the same
$\cos{(\pi p/q_B)}$ dependence on the quasi-momentum $p$, but
different curvatures. Therefore, $m_\mu \ne m_\varepsilon$, with

\begin{eqnarray}
\label{m_e}
{1 \over m_\varepsilon} &\equiv& 
\left.{{\partial^2 \varepsilon }\over {\partial p^2}}\right|_{p=0} 
= {2 \pi^2 ~(K + 2~ \chi~N_0) \over q_B^2} , \\
{1 \over m_\mu} &\equiv&  
\left.{{\partial^2 \mu }\over {\partial p^2}}\right|_{p=0}
 = {2 \pi^2 ~(K + 4~ \chi~N_0) \over q_B^2} ,
\label{m_mu}
\end{eqnarray}
where $K$ and $\chi$ have been defined in Eqs.(\ref{kappa},\ref{chi}).
Sometimes it is convenient to extend the definition of the effective
masses to the full Brillouin zone, introducing the quasi-momentum
dependent masses 

\begin{eqnarray}
&&m_\varepsilon(p) \equiv 
\left[ {\partial^2 \varepsilon  \over \partial p^2} \right]^{-1} 
= {m_\varepsilon \over  \cos(\pi p/q_B)}, \\
&&m_\mu (p) \equiv 
\left[ {{\partial^2 \mu }\over {\partial p^2}}\right]^{-1}  = 
 {m_\mu \over  \cos(\pi p/q_B)},
\end{eqnarray}
where, following Eqs.(\ref{m_e},\ref{m_mu}),
$m_{\varepsilon}=m_{\varepsilon}(0)$ and $m_\mu=m_\mu(0)$.

It is also useful to introduce, with the same line of reasoning, 
two different group velocities, defined as

\begin{eqnarray}
\label{v_e}
{v_\varepsilon} &\equiv&
{{\partial \varepsilon }\over {\partial p}}
= {1 \over m_\varepsilon} {q_B \over \pi} 
\sin \left({ \pi p \over q_B}\right), \\
{v_\mu} &\equiv&
{{\partial \mu }\over {\partial p}}
= {1 \over m_\mu} {q_B \over \pi} \sin \left({ \pi p \over q_B}\right).
\label{v_mu}
\end{eqnarray}
There is a simple, general relation between the two different group
velocities (following from $\mu = \partial~(N_0 \varepsilon) /
\partial N_0$)

\begin{equation}
v_\mu = v_\varepsilon + {{\partial v_\varepsilon}\over {\partial N_0}} N_0
\label{relation}
\end{equation}
with, given Eqs.(\ref{m_e},\ref{m_mu}), $v_\mu > v_\varepsilon$.  The
analog relation for the effective masses has been retrieved in
\cite{menotti1}.

Of course, the concept of {\it effective mass}, defined as the inverse
of the curvature of the corresponding spectrum (as that of {\it group
velocity}, defined as the first derivative) can be extended to shallow
optical potentials, where the nonlinear tight binding approximation
breaks down.  In this case, the quasi-momentum dependence of
$\varepsilon$ and $\mu$ will not be simply described by a cosine
function, but will still remain periodic in the quasi-momentum $p$.  In
particular, the value $p$ where $m_\varepsilon(p)$ changes sign
(corresponding to $\partial ^2 \varepsilon / \partial p^2 =0$) will be
greater than $q_B/2$ and will in general not coincide with the
momentum where $m_\mu(p)$ changes sign (corresponding to $\partial ^2
\mu / \partial p^2 =0$).

The presence of the two {\it different} effective masses (group velocities)
raises an important problem: which effective
mass (group velocity), 
and how, enters in the dynamical properties of the system? For
instance, we anticipate that the current carried by
a Bloch waves with
quasi-momentum $p$ is $\rho_0~ v_\varepsilon (p)$,
where $\rho_0$ is the average particle density;
% and $v_g \equiv \partial \varepsilon / \partial p =
%{q_B \over \pi m_\varepsilon} \sin (\pi p/q_B)$, 
$m_\mu$, on the other hand, plays a
crucial role in the Bogoliubov spectrum. To conclude this subsection,
we notice that the Bloch states are not the only stationary solutions
of the Gross-Pitaevskii equation.  Because of nonlinearity, indeed,
periodic solitonic solutions can also appear for a weak enough periodic
potential, introducing new branches in the excitation
spectra \cite{machholm,li}.

\subsection{Bogoliubov dispersion relation}

In this subsection we study the Bogoliubov spectrum of elementary
excitations. This describes the energy of small perturbations with
quasi-momentum $q$ on top of a macroscopically populated state with
quasi-momentum $p$ (stationary solution of Eq.(\ref{GPE})). To be
explicit, let us consider first the case in which the radial degrees
of freedom $y,z$ are integrated out.  The wavefunction along the $x$
direction can be written as
\begin{eqnarray}
\Psi(x,t)=
e^{-i \mu(p) t / \hbar} e^{ipx/\hbar}
\left[
{\tilde \Psi_p}(x) + \sum_{q}
{\tilde u}_{pq}(x) e^{iqx/\hbar} e^{-i\omega_{pq}t} + 
{\tilde v}^*_{pq}(x) e^{-iqx/\hbar} e^{i\omega_{pq}t}
\right].
\end{eqnarray}
Because of the periodicity, the Bogoliubov amplitudes can be written
as Bloch waves [i.e., $\{u,v\}_{pq}(x)=\exp(iqx/\hbar)
\{\tilde{u},\tilde{v}\}_{pq}(x)$], where $q$ is the quasi-momentum of
the excitation and $\{\tilde{u},\tilde{v}\}_{pq}(x)$ are periodic
functions.
The subscript $\{pq\}$ indicates that both the amplitudes ${\tilde
u},{\tilde v}$ and the excitation frequencies $\omega_{pq}$ depend on
the quasi-momentum $p$ of the carrying wave and on the quasi-momentum
$q$ of the excitation.

In terms of the periodic functions ${\tilde \Psi}$, ${\tilde u}$
and ${\tilde v}$, the Bogoliubov equations take the form 

\begin{eqnarray}
\label{bog_u}
\left[
{1 \over 2m}  \left(-i \hbar \partial_x+p+q \right)^2
+ s \; E_R \; {\rm sin}^2\left({\pi x \over d }\right)
- \mu + 2 g n d |{\tilde \Psi}_p|^2 \right] {\tilde u}_{pq}(x)
&+& g n d  {\tilde \Psi}_p^2 {\tilde v}_{pq}(x) = \cr
&=& \hbar \omega_{pq} {\tilde u}_{pq}(x) \\
\left[
{1 \over 2m}  \left(-i \hbar \partial_x-p+q \right)^2
+ s \;E_R \;  {\rm sin}^2\left({\pi x \over d }\right)
- \mu + 2 g nd |{\tilde \Psi}_p|^2 \right] {\tilde v}_{pq}(x)
&+& g nd {\tilde \Psi}_p^{*2}  {\tilde u}_{pq}(x) = \cr
&=& - \hbar \omega_{pq} {\tilde v}_{pq}(x)
\label{bog_v}
\end{eqnarray}
where $n$ is the 3D-average density and $\int_{-d/2}^{d/2} |{\tilde
\Psi}_p|^2 dx=1$. Equations (\ref{bog_u},\ref{bog_v}) can be solved
numerically in a very efficient way working with the Fourier
components of ${\tilde \Psi}$, ${\tilde u}$ and ${\tilde v}$.

\begin{center}    
\begin{figure}    
\includegraphics[width=0.35\linewidth]{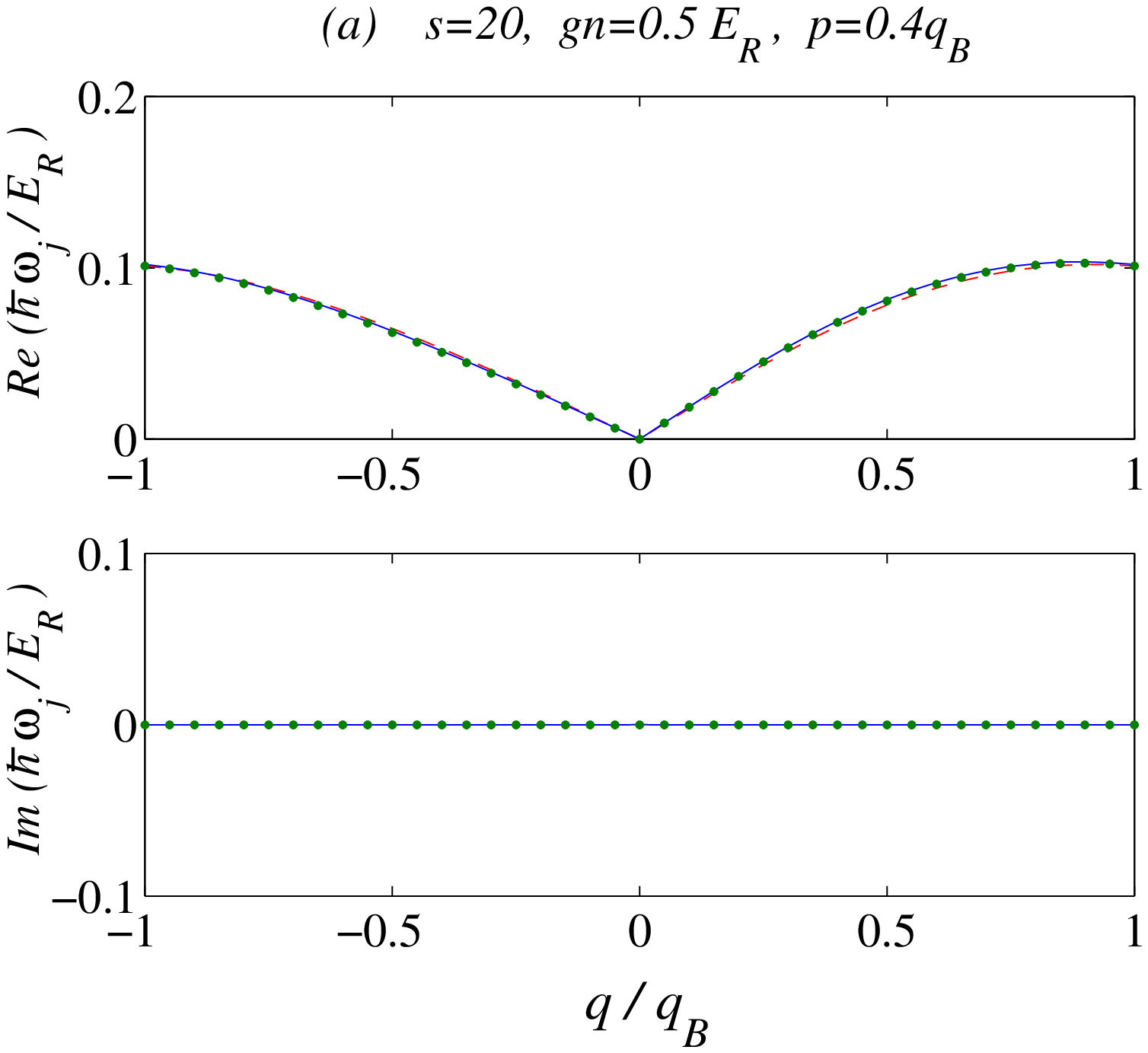}    
\includegraphics[width=0.35\linewidth]{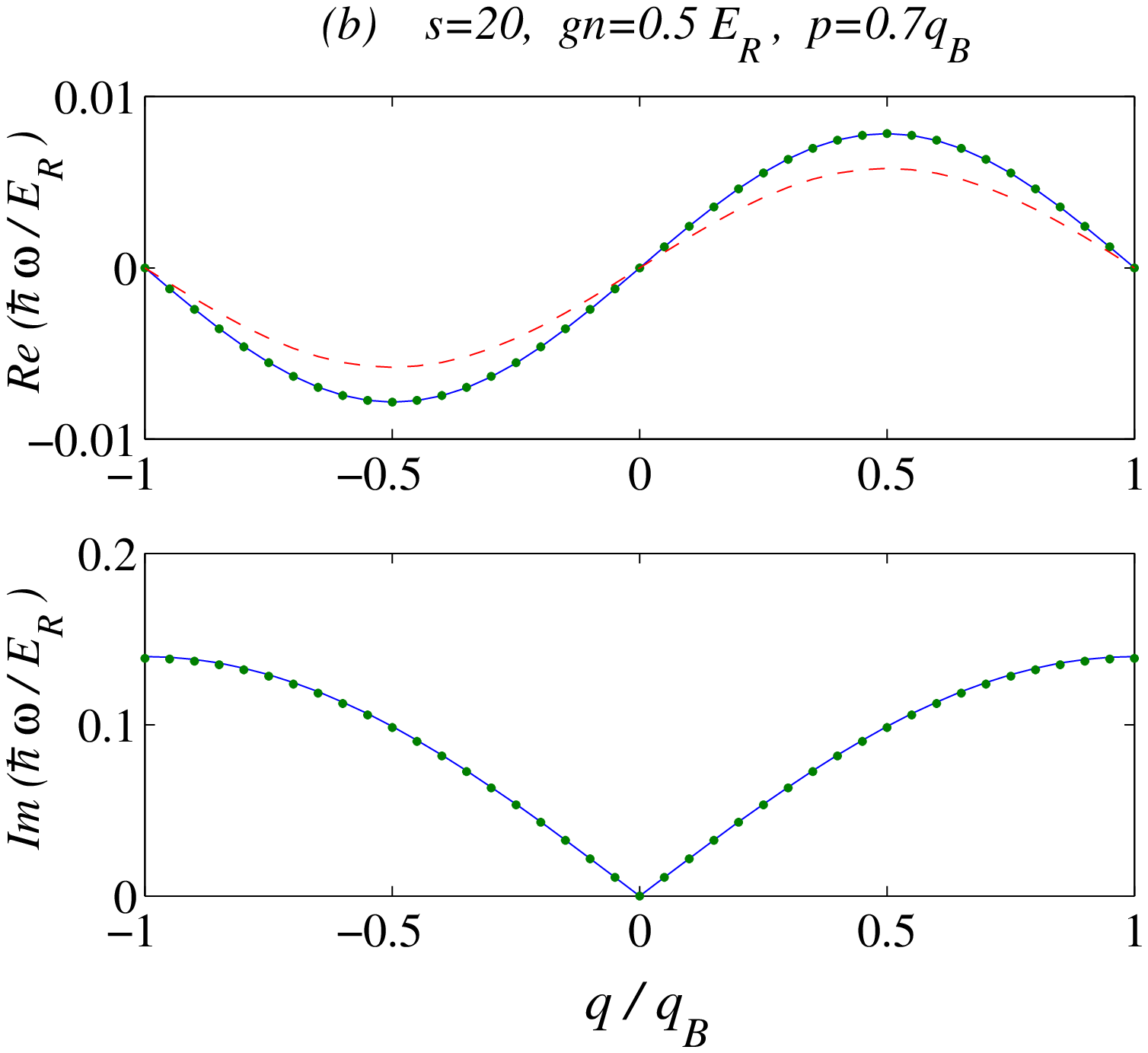}    
\label{check_tb}
\caption{Numerical solutions of Eqs.(\ref{bog_u},\ref{bog_v}) for
$s=20$ and $gn=0.5E_R$ (green dots); analytic solution 
Eq.(\ref{omega}) in the tight binding approximation (solid blue line);
analytic solution of Eq.(\ref{omega}) where $m_\mu$ is replaced
with $m_\varepsilon$ (dashed red line). The quasi-momentum of 
the carrying wave is $p=0.4q_B$ in (a) and $p=0.7q_B$ in (b).}
\end{figure}    
\end{center}    

In the tight binding regime, the Bogoliubov analysis corresponds to
perturbing the large amplitude wave as $\psi_l= [\psi_0+ \delta
\psi_l]~e^{i(p ld-\mu t)/\hbar}$, with $\delta \psi_l = \sum_q {\cal
U}_q e^{i(qld/\hbar-\omega_{pq} t)}$.  Retaining only first order
terms with respect to $\delta \psi$, we get two coupled linear
equations analogous to (\ref{bog_u}) whose eigenvalues can
be calculated analytically \cite{smerzi03-gen}.  The general solution
(for any effective dimensionality of the system: $D=0,1,2,3$) is
\begin{eqnarray}
\label{omega}
&&\hbar \omega_{pq}
= {q_B^2 \over \pi^2 m_\mu} \sin\left({\pi p \over q_B}\right) 
\sin\left({\pi q \over q_B}\right) \pm \nonumber \\
&& \pm 2 \sqrt{ {q_B^4  \over \pi^4 m_\mu^2} 
\cos^2\left({\pi p \over q_B}\right) 
\sin^4\left({\pi q \over 2 q_B}\right) + 
{q_B^2  \over \pi^2 m_\varepsilon} 
\frac{\partial \mu}{\partial N_0} N_0 
\cos\left({\pi p \over q_B}\right) 
\sin^2\left({\pi q \over 2 q_B} \right)} + \nonumber \\
& & + O
\left[\left({{m_\mu^{-1}-m_\varepsilon^{-1}}\over{m_\mu^{-1}}}\right)^2\right]
\end{eqnarray}
with the chemical potential given by $\mu (p,N_0)= \mu^{loc} - {q_B^2
\over \pi^2 m_\mu} \cos \left({ \pi p \over q_B}\right)$ (see
Eq.(\ref{chempot})), and $\mu^{loc}_l = U_{\alpha} \mid \psi_l
\mid^{4/(2+D)}$.  For $D=0$ and in the limit $\chi=0$, we recover the
well-known results for the discrete nonlinear Schr\"odinger equation
\cite{smerzi02,kivshar92}. Equation (\ref{omega}) 
has been first written in \cite{smerzi03-gen} 
in terms of the parameter of the DNL equation Eq.(\ref{DNLS_gen}), while, for
small $q$ and arbitrary $p$, has been derived in \cite{machholm03} 
for arbitrary values of $s$.

In Fig. 1, we compare the analytic results (dots) with the
numerical solution of Eqs.(\ref{bog_u},\ref{bog_v}) (solid line), for
a system in tight binding regime.  In the numerical calculations, the
effective masses have been obtained from the curvatures of the Bloch
energy and chemical potential spectra, while the term $N_0 {{\partial
\mu}\over{\partial N_0}}$ has been evaluated from the density
dependence of the chemical potential. As it has already been noted in
\cite{menotti1}, effects related with the difference between the two
effective masses in the Bogoliubov spectrum of a condensate at rest
($p=0$) are usually negligible. On the contrary, such difference
becomes important when the condensate moves with a large quasi-momentum,
as shown in Fig. 1(b).

\section{SOUND WAVES \& INSTABILITIES}

The small $q$ (large wavelength) limit of the Bogoliubov dispersion
relation becomes

\begin{eqnarray}
\hbar \omega_{pq} \approx 
{q_B \over \pi m_\mu} \sin\left({\pi p \over q_B}\right) \; q
+ |q| \sqrt{{1 \over  m_\varepsilon} \frac{\partial \mu}{\partial N_0} N_0 
\cos\left({\pi p \over q_B}\right) },
\end{eqnarray}
(we assume, for the moment, that
$\frac{1}{m_\varepsilon}\frac{\partial \mu}{\partial N_0} N_0 \cos(\pi
p /q_B) > 0$).  The linear behaviour in $q$ indicates that the system
supports (low amplitude) sound waves, propagating on top of the large
amplitude traveling wave $\Psi_p$ with velocity

\begin{equation}
v_{s,\pm} = \left. \hbar 
\frac{\partial \omega}{\partial q}\right|_{q \to 0^{\pm}} =
\left\{\begin{array}{l l}
v_\mu + c \; ,  \hspace*{0.5cm}  (q \to 0^+) \\
v_\mu - c \; ,  \hspace*{0.5cm}  (q \to 0^-) \\
\end{array}
\right.
\label{sound}
\end{equation}
where the ``chemical potential group velocity'' $v_\mu$ has been defined
in Eq.(\ref{v_mu}), and the ``relative sound velocity'' $c$
is defined as
\begin{equation}
c = \sqrt{ {1 \over m_\varepsilon} 
\frac{\partial \mu}{\partial N_0} N_0 \cos\left({\pi p\over q_B}\right)}.
\end{equation}
The two different velocities $v_{s,\pm}$ correspond, respectively, to a sound
wave propagating in the same and in the opposite direction of the large
amplitude traveling wave. As we have already noticed, and we will 
discuss again in the next section, 
$v_\mu$ is different from (it is larger than) the actual velocity of
the large amplitude wave, see Eq.(\ref{relation}).

We remark that, contrary to the case of a Galilean invariant system
($s = 0$), the sound velocity depends on the quasi-momentum
$p$. Moreover, $v_s$ depends on the effective dimensionality of the
condensates, since (from Eq.(\ref{mu_dimensional},\ref{chempot}))
$\frac{\partial \mu}{\partial N_0} N_0 \sim \alpha~ U_\alpha~
N_0^{\alpha/2}$.  In the limit $\alpha=2$, $p\to 0$ and
$m_\varepsilon,m_\mu \to m$ we get the sound velocity in the uniform
case.

The system is energetically unstable if there exist any
$\omega_{pq}<0$.  In the limit $s = 0$, this corresponds to a group
velocity larger than the sound velocity (Landau criterion for
superfluidity).  When the system has a discrete translational
invariance ($s > 0$) the condition for this instability is obtained
from the Bogoliubov excitation spectrum Eq.(\ref{omega}).  Then, we
have that the system is not superfluid when $\omega_{pq} < 0$,
corresponding to

\begin{equation}
 v_\mu^2  >  c^2 .
\label{ei}
\end{equation}
This result should be compared with the well known Landau criteria for
an homogeneous system ($s=0$), stating that the superfluid is
energetically unstable when $v^2 > c^2$, $v \equiv {{\partial
\varepsilon }\over {\partial p}} = {{\partial \mu }\over {\partial
p}}$ being the group velocity of the condensate, and $c = \sqrt{ {1
\over m} \frac{\partial \mu}{\partial N_0} N_0}$ the sound velocity.

There is a further dynamical (modulational) instability mechanism
associated with the appearance of an imaginary component in the
Bogoliubov frequencies, which disappears in absence of interatomic
interactions, or in the translational invariant limit (if $a > 0$). The
onset of this instability in the tight binding regime, coincides with
the condition

\begin{equation}
c^2 < 0 
\;\; \Rightarrow \;\;
\cos \left( {\pi p \over q_B } \right) < 0 
\;\; \Rightarrow \;\;
|p| > \frac{q_B}{2}.
\label{mi}
\end{equation}
The dynamical instability drives an exponentially fast increase of the
amplitude of the (initially small) fluctuations of the condensate.
Since the initial phases and amplitudes of the fluctuation modes are
essentially random, their growth induce a strong dephasing of the
condensate, and dissipates its translational kinetic energy (which is
transformed in incoherent collective and single particles
excitations).  The unstable modes $q$ grow with a time scale given by
the imaginary part of the excitation frequency

\begin{equation}
\tau_{p q}^{-1} = {2 q_B \over \pi \hbar} 
\left| \sin \left({\pi q \over q_B}\right) \right| 
Im\left[ \sqrt{  {q_B^2 \over \pi^2 m_\mu^2} 
\cos^2\left({\pi p \over q_B}\right) 
\sin^2\left({\pi q \over 2 q_B}\right) + 
{1 \over m_\varepsilon} \frac{\partial \mu}{\partial N_0} N_0 
\cos\left({\pi p \over q_B}\right) } \right]. 
\label{t_mi}
\end{equation} 
We remark here the different scaling of the energetic and dynamical
instability with the interatomic interactions. Decreasing the
scattering length, the sound velocity decreases, and smaller and
smaller group velocities can breakdown the superfluidity of the system
(in the limit $a=0$, the sound velocity $c=0$: the non interacting
condensate is {\rm always} energetically unstable for an arbitrary
small group velocity). On the other hand, the dynamical modulational
instability criteria does not depend on the scattering length. This
apparent paradox is simply solved noticing that the growth time of the
unstable modes, Eq.(\ref{t_mi}), actually depends on interactions, and
diverge when the scattering length vanishes ($\tau \to \infty$ when $a
\to 0$).  Therefore, a noninteracting condensate is always dynamically
stable.  There is a further point to remark: if we consider a
condensate moving with an increasing velocity, the system always
becomes first energetically unstable, then it hits the dynamical
instability. As a matter of fact, however, in real experiments the
energetic instability can grow quite slowly (and at zero temperature
only in presence of impurities \cite{wu01}), so that the dominant
dephasing mechanism is given by the modulational instability. This
aspect can be highlighted also with numerical experiments, studying,
for instance, Bloch oscillations of a condensate with the interactions
switched off. In this case, even though the system is energetically
unstable, it remains coherent over many oscillations. If the
interatomic interactions are switched on, however, the system dephase
rather quickly, the dephasing occurring when the quasi-momentum of the
condensate is in the dynamical unstable region of the Bloch
spectrum. We have done such numerical experiment, and results are
shown in Movies 4 and 5. Of course, our prediction can be tested in real
experiments, tuning the scattering length with a Feshbach resonance.

To summarize, the tight binding approximation predicts the arising of
the dynamical instability (complex excitation frequency) for
$p=q_B/2$.  We point out that $p=q_B/2$ also corresponds to the
quasi-momentum where, in the tight binding regime, the effective
masses $m_\varepsilon(p)$ and $m_\mu(p)$ change sign. A system with a
negative effective mass and positive scattering length can be, roughly
speaking, seen as equivalent to a system with a negative scattering
length and positive effective mass.  It is well known that a BEC
having a negative scattering length is dynamically unstable, and,
therefore, such parallelism could be proposed to give a simple
explanation of the instability. However, we will see that this
coincidence between the arising of dynamical instabilities and the
inversion of sign of the effective mass does not take place at
lower optical potentials (see Fig. \ref{m_star_m_im}).

\begin{center}    
\begin{figure}    
\includegraphics[width=0.3\linewidth]{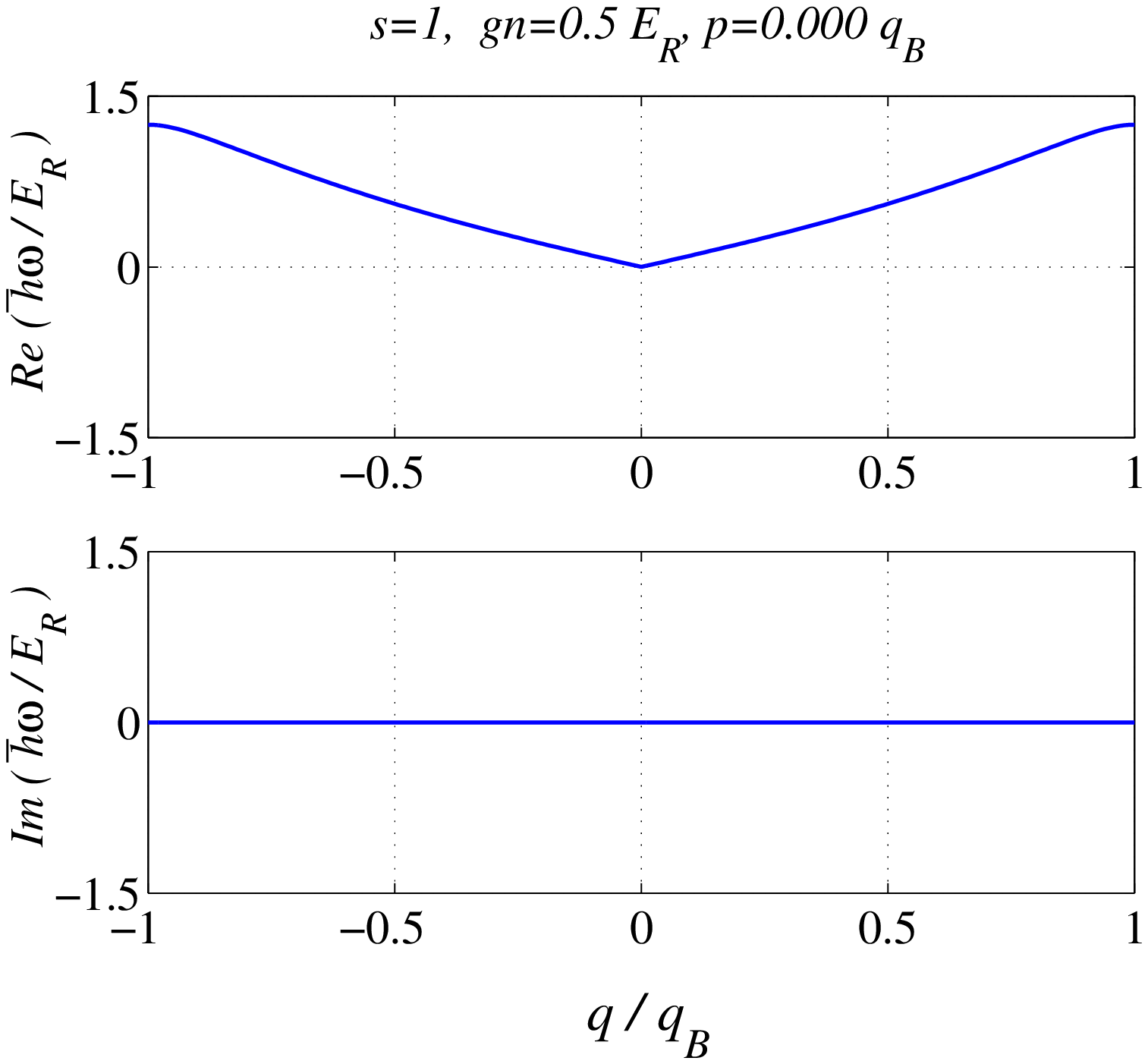}    
\includegraphics[width=0.3\linewidth]{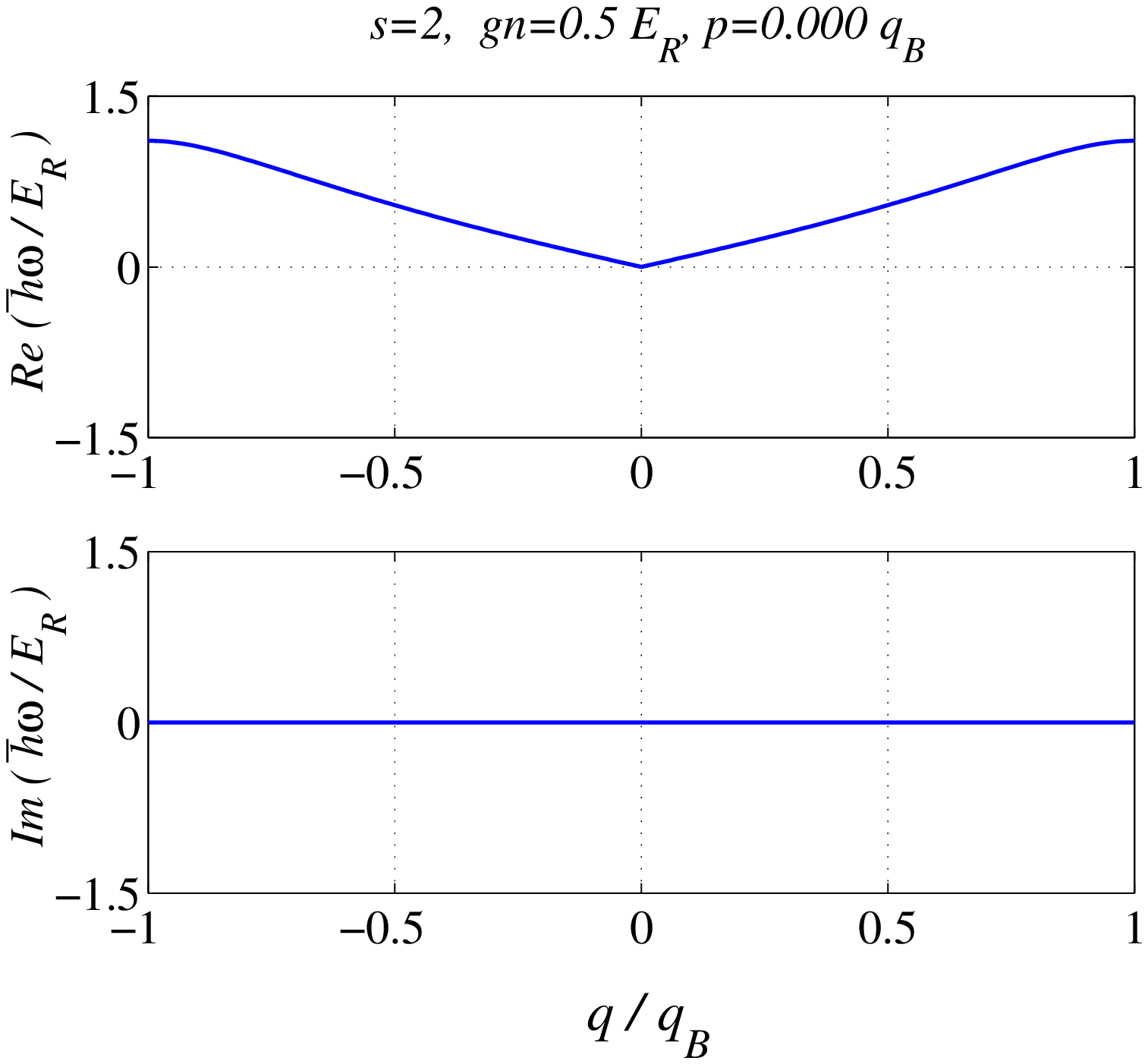}    
\includegraphics[width=0.3\linewidth]{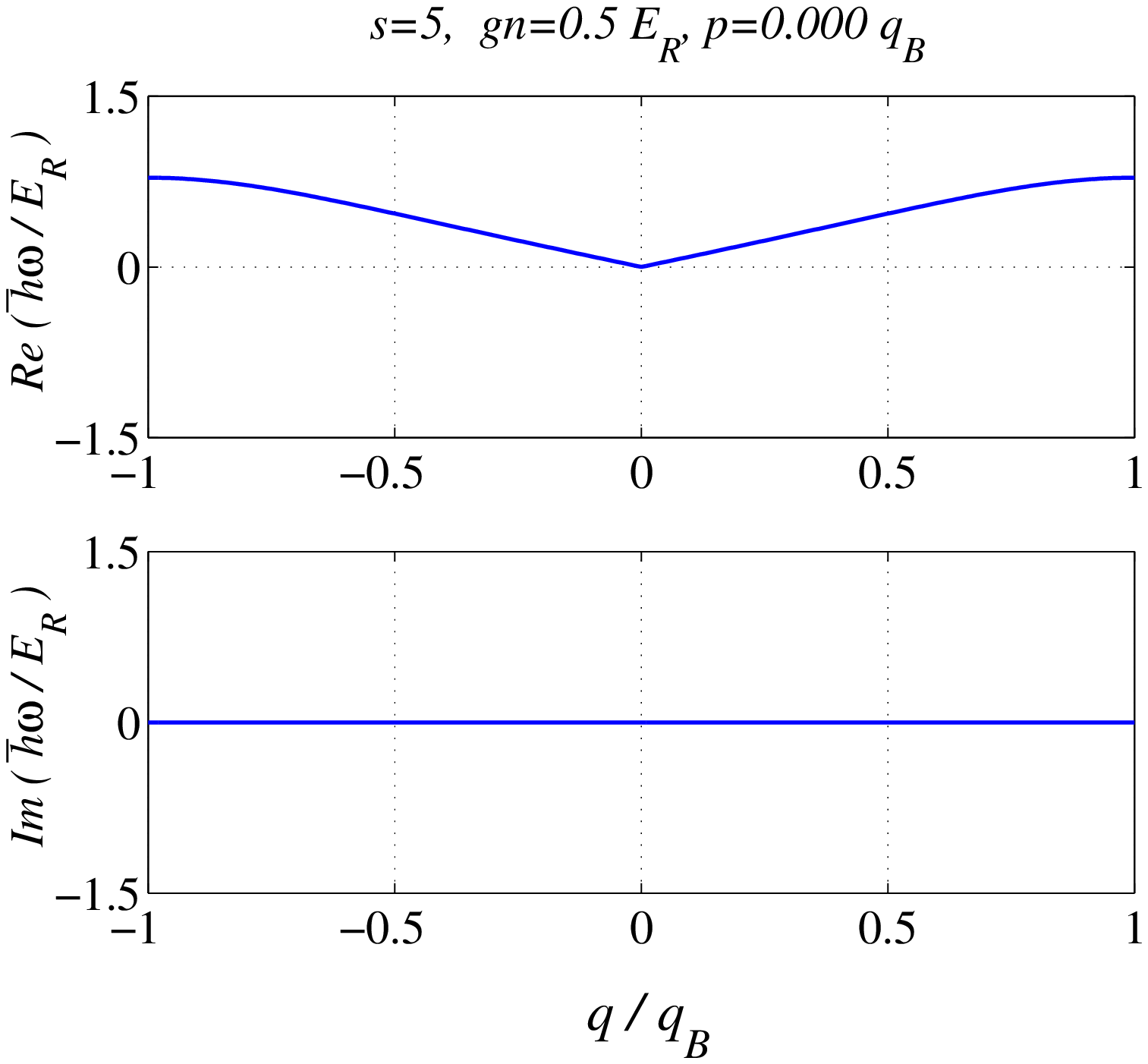}    
\caption{MOVIES: Real and imaginary part of $\omega_{pq}$ as a function of $q$
for different values $p$, for $gn=0.5E_R$ and $s=1$ (a), $s=2$ (b) and
$s=5$ (c).}
\label{movie_omega}
\end{figure}    
\end{center}

\begin{center}    
\begin{figure}    
\includegraphics[width=0.35\linewidth]{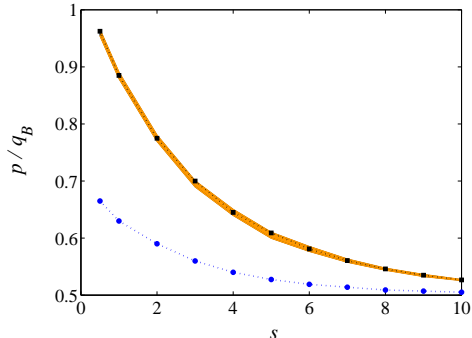}    
\caption{Results from the numerical solution of
Eqs.(\ref{bog_u},\ref{bog_v}) for $gn=0.5 E_R$ as a function of
$s$. Dots: value of the quasi-momentum $p$ where the excitation
frequencies $\omega_{pq}(q=\pm q_B)$ become complex; orange region:
value of the quasi-momentum $p$ where complex frequencies are found
around $q=0$; squares: value of the quasi-momentum $p$ where the
effective mass $m_\varepsilon$ changes sign. The dotted lines are a
guide to the eye. }
\label{m_star_m_im}
\end{figure}    
\end{center}

Let us concentrate now on the behaviour of the excitation frequencies
for shallow optical potentials, where the tight binding expression
derived in (\ref{omega}) is not applicable.  For small optical
potential depths, the Bogoliubov equations have to be solved
numerically and the results show a more complicated behaviour.  In
Movies \ref{movie_omega}(a-c), we show the numerical solutions of the
1D Bogoliubov equations for three different values of $s$ ($s=1, 2$
and $5$). In those movies, we plot the real and imaginary part of
$\omega_{pq}$ as a function of $q$ and we vary $p$ in time.

We point out a series of differences with respect to the tight binding
regime:

- the complex frequencies appear at the boundary of the first
Brillouin zone ($q=q_B$) for a value of $p>q_B/2$ (dots in
Fig. \ref{m_star_m_im}) and they reach the center of the zone ($q=0$)
for a higher value of $p$ (orange region in Fig. \ref{m_star_m_im}).

- the range of momenta $p$ where the frequencies are complex for some
$q$, but real around $q=0$, decrease by increasing $s$. In the tight
binding limit this range vanishes;

- in the limit of our numerical accuracy, which is due to the discrete
sampling of $p$ and $q$, we found that the value of $p$ where the
effective mass $m_\varepsilon$ changes sign (squares in
Fig. \ref{m_star_m_im}) corresponds to the value of $p$ at which the
frequencies with non vanishing imaginary part reach $q \approx 0$. In
the tight binding approximation, this appears explicitly through the
term $\cos(\pi p /q_B)/m_\varepsilon$ under the square root.

We would like to remark two important results arising from our study
of the excitation spectra.  First, as shown in Fig.\ref{m_star_m_im},
we found the onset of the dynamical instability for values of the
quasi-momentum where the effective mass $m_\varepsilon(p)$ is still
positive. Second, the range of momenta where the system has a positive
effective mass and, at the same time, is dynamically unstable,
increases by decreasing $s$ (keeping in mind that the amplitude of the
imaginary part vanishes for $s \to 0$ or $gn \to 0$, which implies
that the growth in time of the instability diverges both for uniform
interacting systems and for an ideal gases in optical lattices). So,
one can study the behaviour of the system at low $s$ to distinguish
between two possible dephasing mechanisms, one due to the sign of
$m_\varepsilon$, the other one due to the dynamical instability, as it
will be extensively explained in Sect.\ref{exps}.

We conclude this section remarking that various important aspects of
the physics of energetic and dynamical instabilities of a BEC in a
periodic potential have been studied in
\cite{trombettoni01,wu01,smerzi02,machholm03}.

%%%%%%%%%%%%%%%%%%%%%%%%%%%%%%%%%%%%%%%%%%%%%%%%%%%%%%%%%%%%

\section{NEWTONIAN DYNAMICS} 
\label{newton}

Using the results in \cite{smerzi03-gen}, we can now rewrite the
dynamics of a BEC wave-packet in terms of the energy effective
mass. For the BEC wave-packet we use the following ansatz

\begin{equation}
\psi_l=\sqrt{\cal K(\sigma)} f\bigg(\frac{l-\xi}{\sigma}\bigg) 
e^{iP(l-\xi)+i \frac{\delta}{2}(l-\xi)^2} , 
\label{var}
\end{equation}  
where $\xi(t)$ and $\sigma(t)$ are, respectively, the center and the
width (in lattice units) of the wavepacket, $P(t)$ and $\delta(t)$
their associated momenta and $\cal K(\sigma)$ a normalization factor
such that $\sum_l N_l=N_T$ (with $ |\psi_l|^2 \equiv N_l$). The
function $f$ is generic, for instance, we can choose $f(X)=e^{-X^2}$
or $f(X)=(1-X^2)^{1/\alpha}$ (with $-1 \le X \le 1$) to describe,
respectively, the dynamics of a gaussian or a Thomas-Fermi wavepacket.
The equations of motion of the collective variables
$\xi(t),\sigma(t),p(t),\delta(t)$ have been obtained in
\cite{trombettoni01,smerzi03-gen}. With ${\cal V}_j=\Omega j^2$
($\Omega=m d^2 \omega_x^2/2$), and neglecting the dynamics wave-packet
width dynamics ($\dot{\sigma}(t) = 0$), we find that the group
velocity $\dot{\xi}$ and the effective force acting on the center of
mass of the wavepacket are given by

\begin{eqnarray}
\label{group}
\hbar \dot{\xi} &=& 
\frac{q_B^2}{\pi^2} \left\langle {1\over  m_\varepsilon} \right\rangle 
\sin{P}, \\ 
\hbar \dot{P} &=& - \frac {\partial V_d}{\partial \xi},
\end{eqnarray}
where $V_d=\Omega (\xi^2+\sigma^2 \frac{ {\cal I}_2 } { {\cal I}_1 }
)$ with ${\cal I}_1=\int dX f^2(X)$ and ${\cal I}_2=\int dX X^2
f^2(X)$.  Since the effective masses depend on the local (on-site)
density, we have to introduce an effective mass averaged over the
local density of the condensate wavepacket

\begin{equation}
\left\langle{1 \over m_\varepsilon}\right\rangle =
{ {\sum_l m_\varepsilon^{-1}(N_l) |\psi_l|^2} \over {\sum_l |\psi_l|^2} },
\label{m_av}
\end{equation}
with, according to Eq.(\ref{m_e}), $m_\varepsilon^{-1}(N_l)=(2
\pi^2/q_B^2) (K+2\chi N_l)$.  We summarize here the most important
results, written in term of the effective mass $m_\varepsilon$:

$(i)$ in the case of an homogeneous system ($V_D =0,~N_l = const.$),
the tunneling rate is given by

\begin{equation}
\frac{\pi^2 \hbar}{q_B^2}
{  \dot{N}_l^{out} \over { N_l \Delta \phi}}|_{\Delta \phi \to 0} =
{1 \over m_\varepsilon};
\end{equation}

$(ii)$ the frequency of small amplitude oscillations of the wavepacket
driven by an harmonic field ${\cal V}_l \propto l^2$ is

\begin{equation}
{ \omega_{dip} \over \omega_x} = 
\sqrt{ \left\langle {m \over m_\varepsilon} \right\rangle } ;
\label{omega-dip}
\end{equation}

$(iii)$ if the driving field is linear ${\cal V}_l=m G d l $, we have simple 
Bloch oscillations with

\begin{equation}
\xi = \left\langle {1 \over m_\varepsilon} \right\rangle 
{q_B^2 \over {\pi^2\,m \,G d}} \cos \left({\pi m \,G \over q_B} t\right) 
\;\;\;.
\label{bo}
\end{equation}
This analysis does not take into account possible dephasing mechanisms
as those investigated in the previous section. In the collective coordinate
approach, such dephasing mechanisms can be described including the dynamics
of the width of the wavepacket $\sigma(t)$ and of the corresponding
momentum $\delta(t)$ \cite{trombettoni01}.

\section{NUMERICAL EXPERIMENTS ON BLOCH OSCILLATIONS, DIPOLE OSCILLATIONS 
AND FREE EXPANSIONS IN THE LATTICE}
\label{exps}

In this section, we discuss some numerical simulations of the
Gross-Pitaevskii equation in order to illustrate the phenomena
described in the previous sections.  We first consider Bloch
oscillations (Sect.\ref{s_bloch}): we create the condensate in a
harmonic trap superimposed to the lattice and then switch off the
harmonic trap and replace it with a linear potential. We expect that
the BEC oscillates periodically in space (Bloch oscillations).

The second numerical simulation consists in creating the condensate in
a harmonic trap superimposed to the lattice, and suddenly displacing
the center of the harmonic trap (Sect.\ref{s_dipole}). This experiment
has already been studied theoretically \cite{smerzi02} and performed
experimentally in \cite{cataliotti01,cataliotti03}. We discuss it
again, generalizing the previous results to the case of shallow
optical lattices.

The third numerical simulation consists in creating a condensate in a
harmonic trap superimposed to the optical lattice and, thereafter,
switching off the harmonic trap in the lattice direction, letting the
condensate expand in the periodic potential (Sect.\ref{s_expansion}):
for values of the the mean-field energy large enough (with a fixed
height $s$ of the interwell energy barriers), the wave-packet is
self-trapped \cite{trombettoni01} and the spreading of the wave packet
does not occur.

In all cases, for an interacting BEC, we have found some sort of
self-trapping and dynamical instabilities for some values of the depth
of the periodic potential or the initial conditions of the BEC wave
packet.  For instance, in the dipole oscillation experiment, the
condensate may stop on one side of the harmonic potential being unable
to complete the oscillation.  It is useful to look at the dynamical
evolution of the relative phases of condensates trapped in neighboring
wells and the BEC evolution in momentum space. There is a clear
correspondence between the distribution in {\it momentum} space and
that in {\it quasi-momentum} space: the quasi-momentum distribution of
the Bloch state $\Psi_p(x)=e^{ipx/\hbar} {\tilde \Psi_p}(x)$ with
quasi-momentum $p$ is $\delta(p)$; its momentum distribution is $|
\sum_{\ell} c_{\ell} \delta(p+2\ell q_B) |^2$, where $\ell$ are
integers and where the $c_{\ell}$ are the Fourier coefficient of the
periodic function $ {\tilde \Psi_p}(x)$. An analogous relation is also
valid when the condensate is not in a well-defined Bloch state, but in
a superposition of Bloch states of the first band. In this case, the
width of the peaks of the momentum distribution will be equal to the
width of the quasi-momentum distribution. In the following we will
work with the momentum distribution, which is simply obtained from the
Fourier transform of the condensate wavefunction.

\subsection{Bloch oscillations}
\label{s_bloch}

Bloch oscillations can be explained in very simple terms.  In the
presence of a linear potential superimposed to the optical lattice,
the behaviour of a particle with quasi-momentum $p$ is described by
the equation of motion $p(t)=F t$, where $F$ is the constant force due
to the linear potential. Since the velocity is given by $v_g=\partial
\varepsilon(p) / \partial p$, when the effective mass is negative, the
particle will respond to a positive (negative) force with a negative
(positive) acceleration. Since the energy band $\varepsilon (p)$ is
periodic in $p$, this will result in periodic oscillations in
coordinate and velocity space.

\begin{center}    
\begin{figure}    
\includegraphics[width=0.35\linewidth]{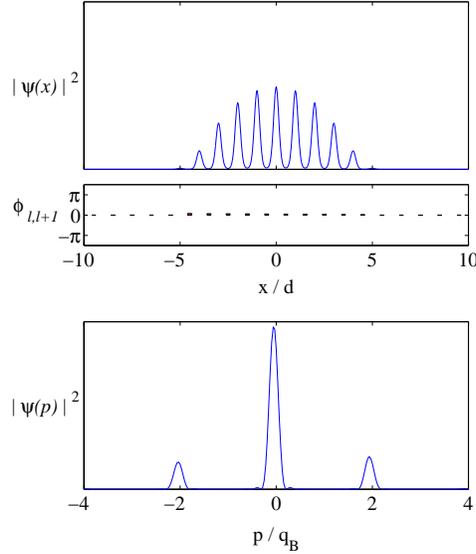}    
\caption{MOVIE: 
Bloch oscillations for $s=10$ and $gn=0.5 E_R$. Time evolution of
the spatial density (upper plot), of the relative phases (middle plot)
and of the momentum distribution (lower plot).}
\label{movie_bloch_1}
\end{figure}    
\end{center}

This simple explanation, even neglecting important effects like
Landau-Zener tunneling to a higher band, 
provides a useful model to interpret experiments with electrons
\cite{electrons}, with cold atoms \cite{dahan96} and with
Bose-Einstein condensates \cite{anderson98,morsch01}.  However, if
interactions in the condensate play a mayor role, the scenario can
change dramatically. First of all, the momentum distribution
$\Psi(p,t)$ will not evolve just like $\Psi(p(t))$, as it
approximately happens for non interacting systems, but will also
spread.  Furthermore, in the presence of interactions, it might happen
that the condensate gets dephased and, after a short while, the
oscillations stop (see Movie \ref{movie_bloch_1}).  For the situation
described in the movie ($s=10$, for which the tight binding
approximation works well), the dephasing process begins when the
center of the momentum distribution reaches $p=q_B/2$.  This point
corresponds both to the on-set of the dynamical instabilities and the
inversion of the effective mass. Since the momentum distribution has a
certain width, one could think that the oscillations stop because the
sign of the effective mass is not the same for the whole
condensate. An alternative interpretation relies on the onset of
dynamical instabilities.

\begin{center}    
\begin{figure}    
\includegraphics[width=0.35\linewidth]{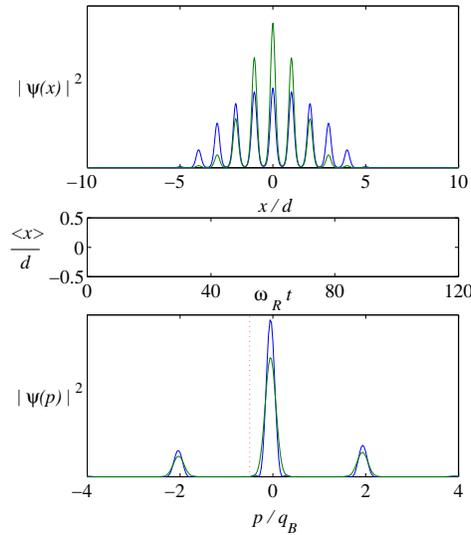}    
\caption{MOVIE:
Bloch oscillations with $s=10$, $gn=0$ (green line) and
$gn=0.5 E_R$ (blue line). Time evolution of the spatial density (upper
plot), of the center of mass (middle plot) and of the momentum
distribution (lower plot). The red dotted line indicates the onset
of dynamical instabilities in the interacting case.}
\label{movie_bloch_2}
\end{figure}    
\end{center}

In order to highlight the correct interpretation, we study the Bloch
dynamics of a non interacting condensate, which is always dynamically
stable.  The initial spatial width is chosen in order to get about the
same momentum distribution as in the interacting case, in order to
have similar effective mass effects.  More specifically, since in the
interacting case the width of the momentum distribution increases
slowly, while in the non interacting case it remains almost constant,
we choose the initial conditions so that the two momentum
distributions will be similar at the ``critical point'', where
$\langle p \rangle=q_B/2$.

The direct comparison is shown in Movie \ref{movie_bloch_2}. We
observe, in the non-interacting case, regular, perfectly periodic
Bloch oscillations, in spite of the finite width of the momentum
distribution.  This clearly shows that, in the interacting case, the
onset of decoherence is due to the dynamical instability.

\subsection{Dipole oscillations}
\label{s_dipole}

Dipole oscillations consist in the motion of the condensate at the
bottom of the harmonic trap.  The average velocity is periodic in time
and the momentum distribution, showing the characteristic peaks due to
the optical lattice, also oscillates periodically in time at the
bottom of the band. During the time evolution, the phase differences
between neighbouring condensates remain locked over the whole
condensate.

\begin{center}    
\begin{figure}    
\includegraphics[width=0.24\linewidth]{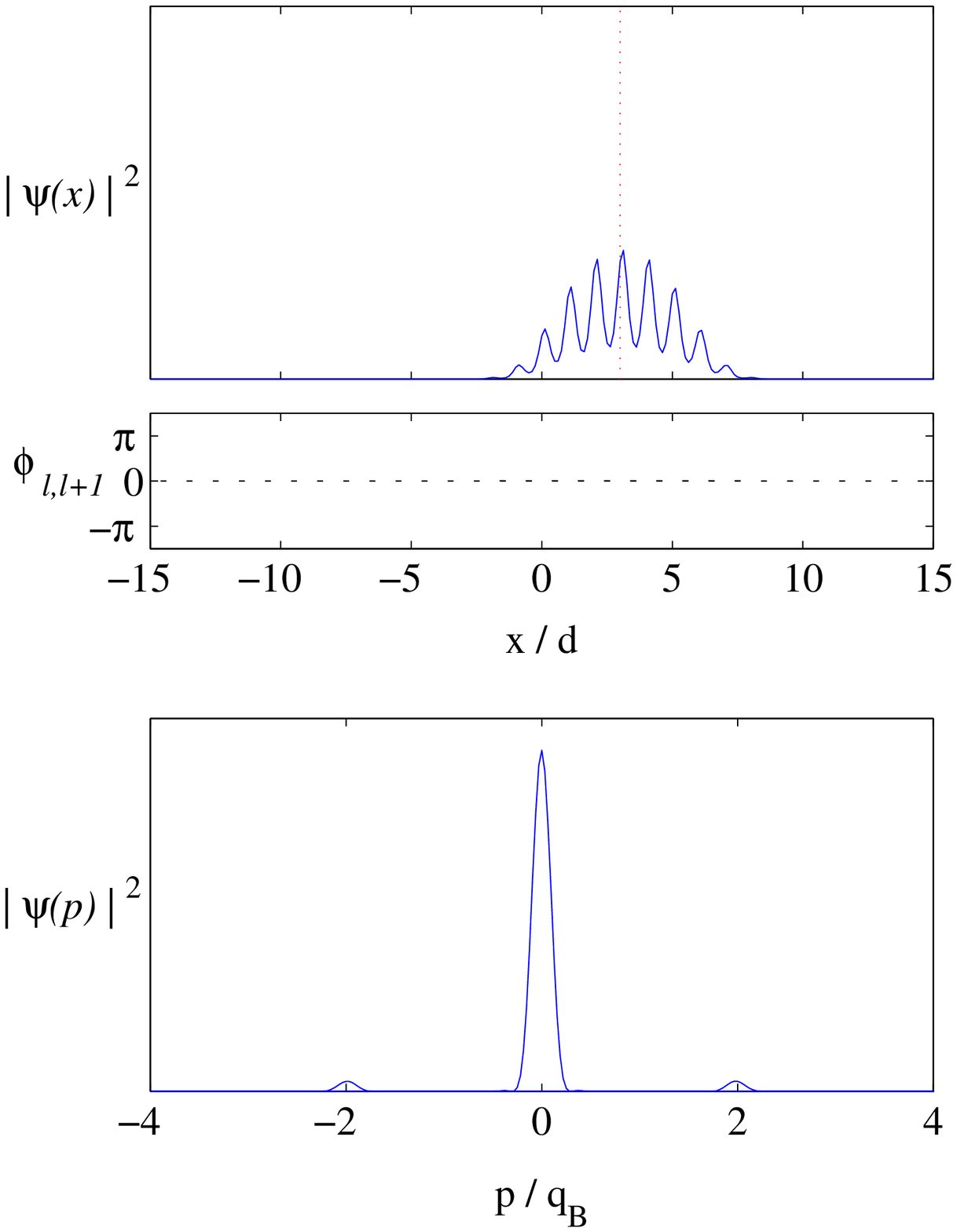}    
\includegraphics[width=0.24\linewidth]{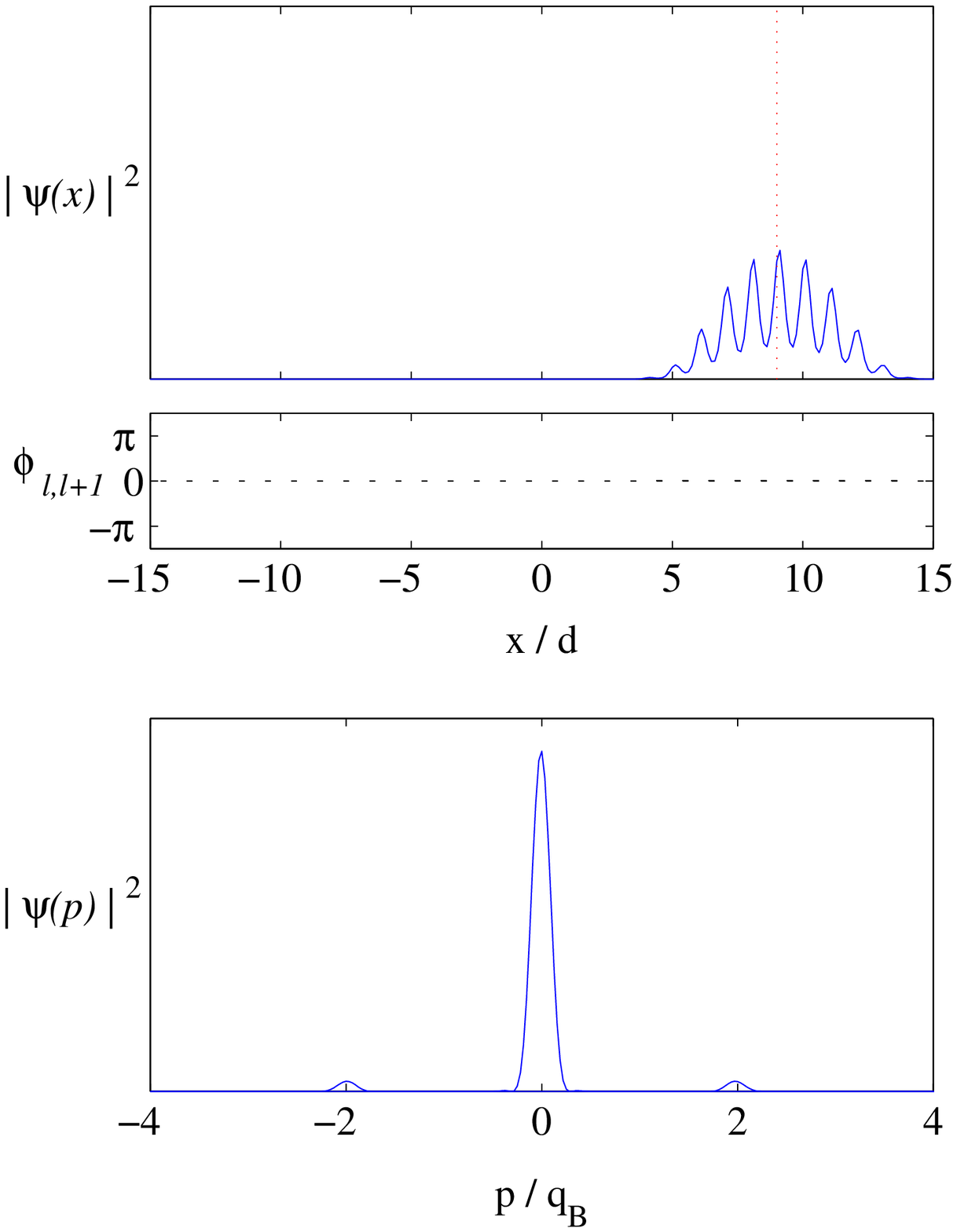}    
\includegraphics[width=0.24\linewidth]{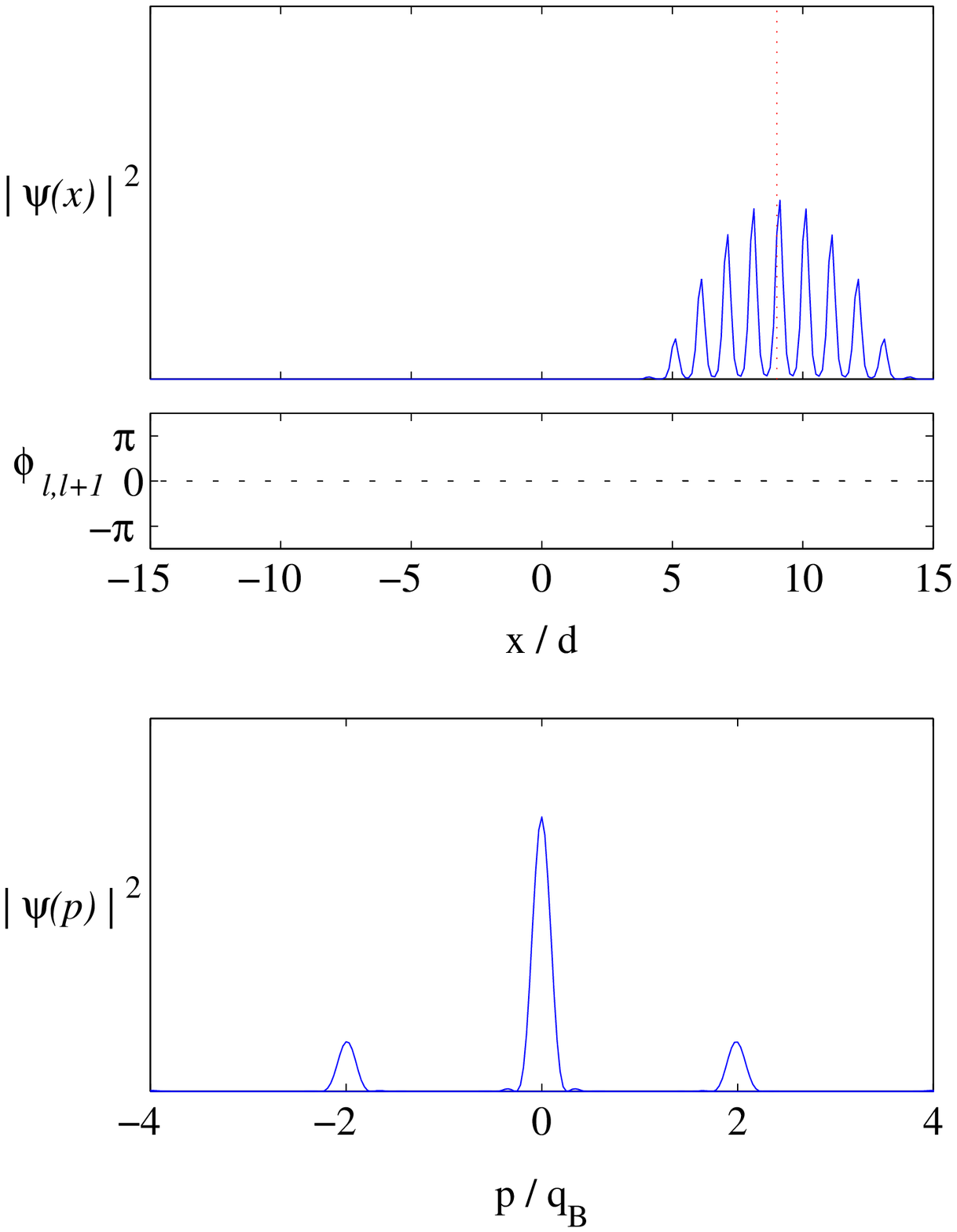}    
\includegraphics[width=0.24\linewidth]{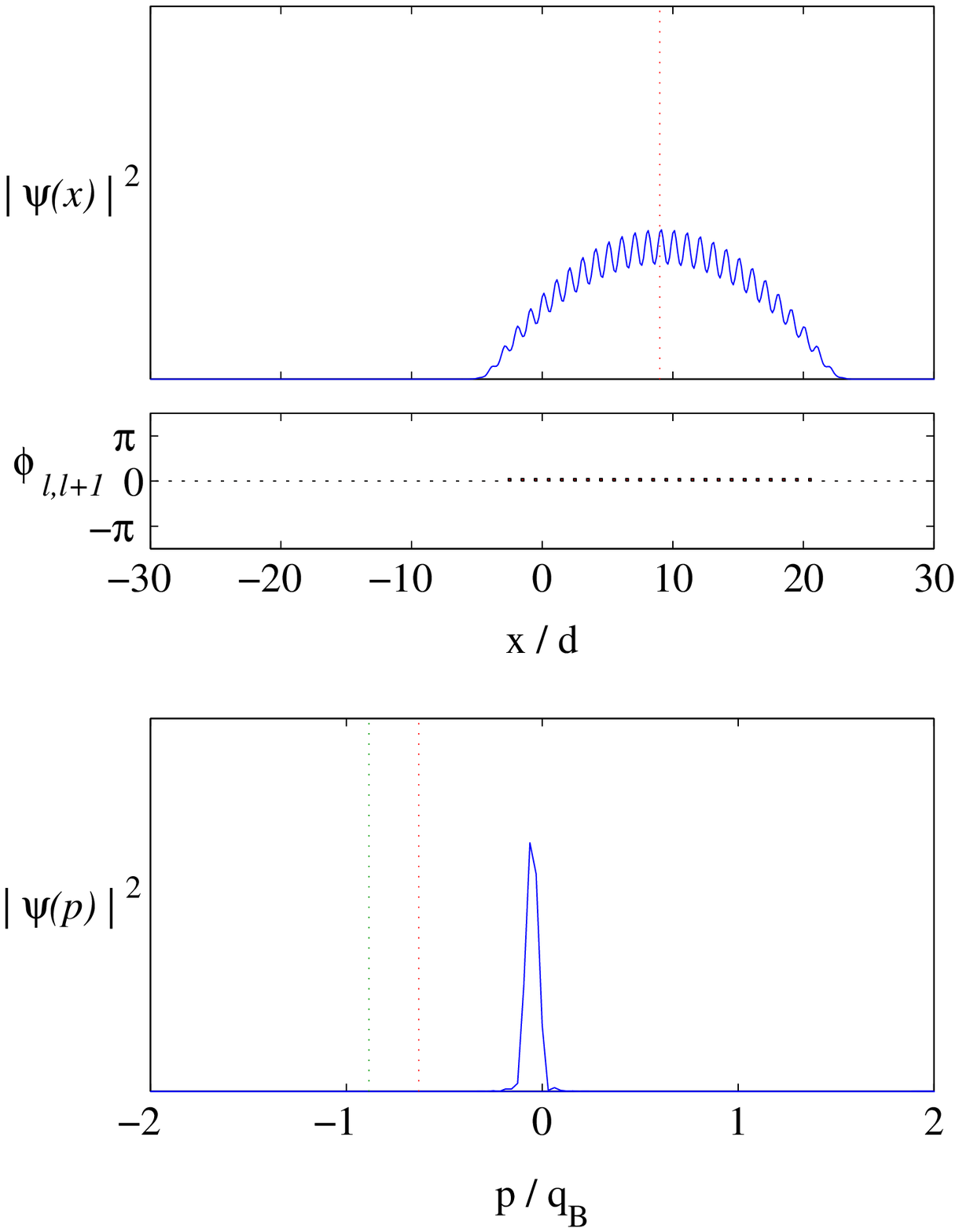}    
\caption{MOVIES: Dipole oscillations for 
$gn=0.01E_R$, $s=3$, $x_0=3d$ (a);
$gn=0.01E_R$, $s=3$, $x_0=9d$ (b);
$gn=0.01E_R$, $s=10$, $x_0=9d$ (c);
$gn=0.5E_R$, $s=1$, $x_0=9d$ (d).
Time evolution of the spatial density (upper plot), of the
relative phases (middle plot) and of the momentum
distribution (lower plot). In movie (d),  we indicate with a red
dotted line the quasi-momentum where the dynamical instabilities arise
and with a green dotted line the quasi-momentum where the effective
mass changes sign.}
\label{movie_dipole}
\end{figure}    
\end{center}

For a given set of parameters corresponding to small displacements,
small interactions or small optical potential depths, dipole
oscillations remain periodic, with the condensate locked in
phase. Instead, increasing one of these quantities, we find that the
oscillations get dephased during the time evolution, or even
stop before the condensate reaches the bottom of the harmonic
potential.
For the seek of comparison we display in Movie \ref{movie_dipole}(a-c)
the evolution of the density, of the phase difference and of the
momentum distribution for the following sets of parameters.  For fixed
interactions ($gn=0.01E_R$) and harmonic trap ($\hbar \omega_x=0.004
E_R$), we choose:

(a) $s=3$, $x_0=3d$: oscillations;

(b) $s=3$, $x_0=9d$: broken oscillations;

(c) $s=10$, $x_0=9d$: broken oscillations.

Looking at the phase difference between neighbouring condensates, we
find that when the condensate oscillation is interrupted,
the phases get scrambled.
This corresponds to a randomized flux of atoms which are not able
anymore to flow coherently downwards the potential.  The evolution of
the momentum distribution suggests that this phenomenon happens when 
the condensate reaches the instability region, given in the specific cases
by $p$ greater than $q_B/2$.

To further explore this interpretation, we choose a shallow optical
potential such that there is a broad range of $p$ where the effective
mass $m_\varepsilon(p)$ is positive and at the same time the system is
dynamically unstable (see Fig. \ref{m_star_m_im}). We increase the
nonlinear interaction parameter to get a relevant imaginary part of
the excitation frequencies, otherwise the time scale where
instabilities manifest themselves is too long.
In Movie \ref{movie_dipole}(d) (lower plot), we indicate with a red
dotted line the quasi-momentum where the dynamical instabilities arise
and with a green dotted line the quasi-momentum where the effective
mass changes sign.
We actually observe the first signatures of decoherence when the
momentum distribution is contained between the two lines, indicating
than the decoherence happens in correspondence of the dynamical
instability point. We conclude this section mentioning that
experimental evidences of dynamical instabilities are reported in
\cite{cataliotti03}.

\subsection{Expansion in the lattice} 
\label{s_expansion}

After creating the condensate in the harmonic trap superimposed to the
lattice, we switch off the harmonic trap and let the condensate, which
is initially at rest, expand. During the expansion, the current of
atoms is from the inside to the outside of the cloud and the phase
differences increase, being positive for $x<0$ and negative for
$x>0$. In \cite{trombettoni01} the occurrence of self-trapping has been
predicted in the tight-binding: when interactions are larger than a
critical value, the width of the wave packet does not continue to
increase with time (as for vanishing or small interactions) and the
wave packet remains localized around a few sites. A similar nonlinear
self-trapping occurs in the two-site problem \cite{smerzi97}.

\begin{center}    
\begin{figure}    
\includegraphics[width=0.35\linewidth]{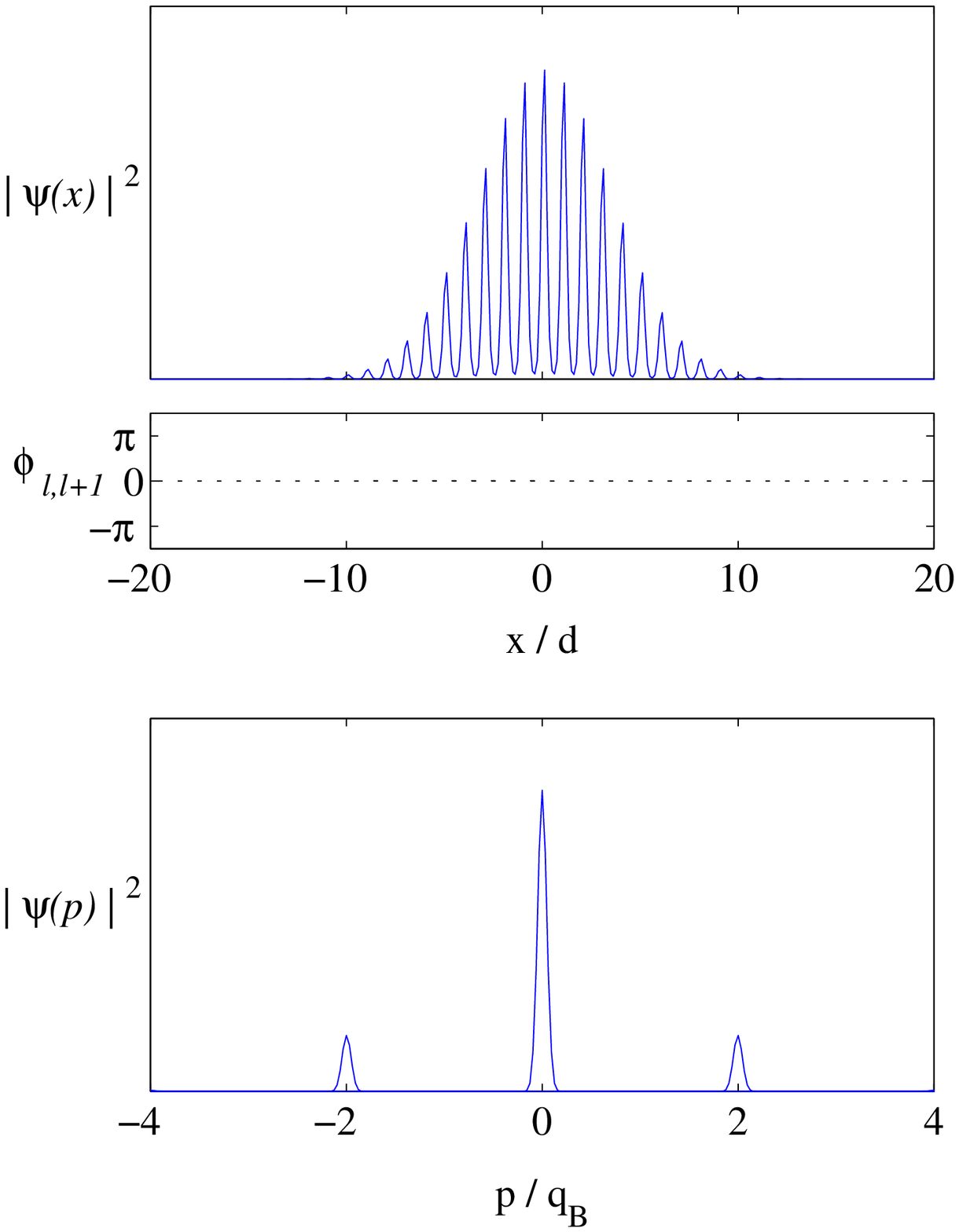}    
\includegraphics[width=0.35\linewidth]{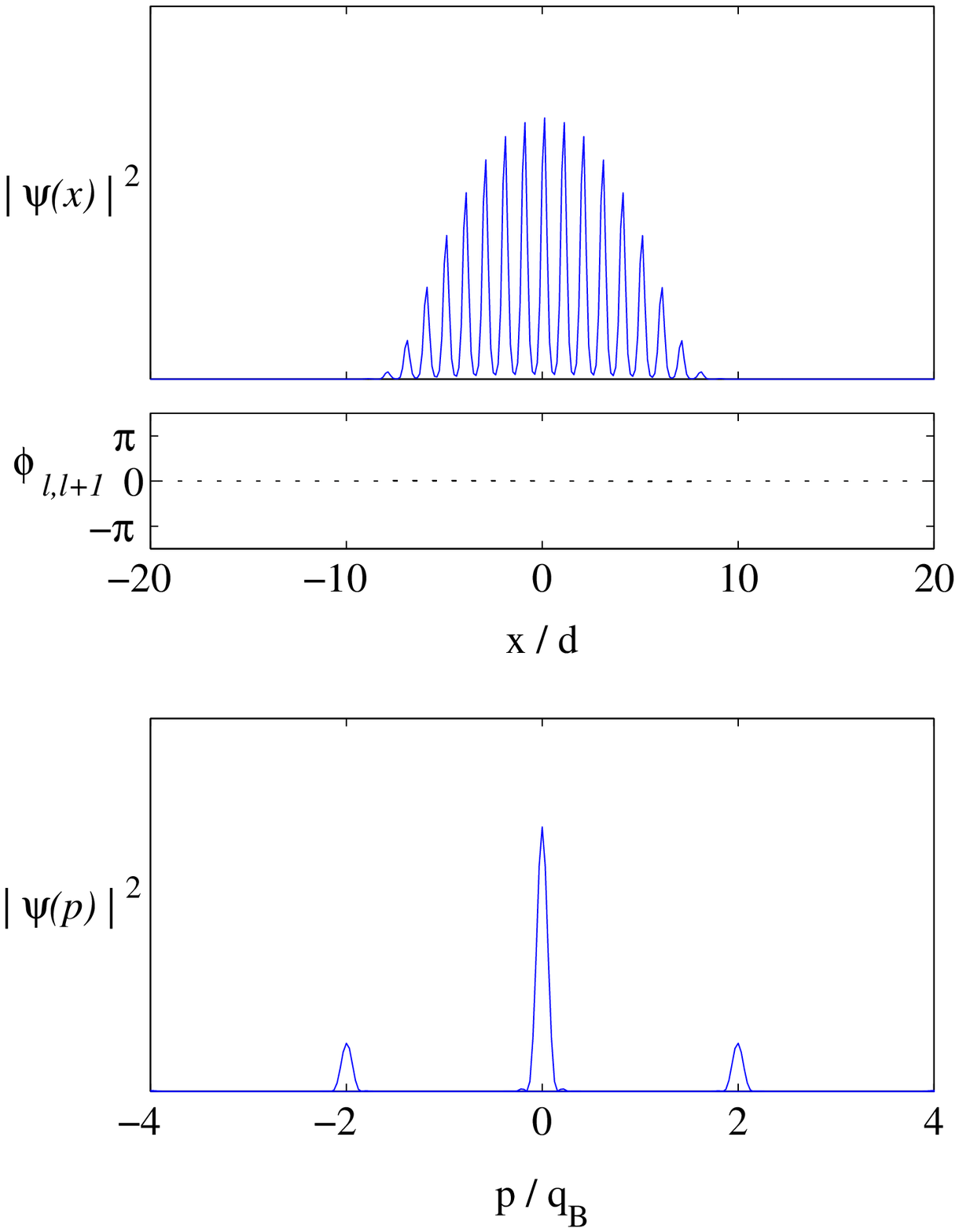}    
\caption{MOVIES: Expansion of a BEC wave-packet for $gn=0$, $s=5$ (a);
$gn=0.01 E_R$, $s=5$ (b);
Time evolution of the spatial density (upper plot), of the
relative phases (middle plot) and of the momentum
distribution (lower plot).}
\label{movie_expansion}
\end{figure}    
\end{center}

Increasing the interactions, the system enters in the self-trapped
regime as shown in Movie \ref{movie_expansion}(b). If interactions are
strong enough, we see that after a first stage (whose duration depends
on the strength of the interactions) the expansion stops and the
condensate evolves as a random flow of atoms between the condensates
localized at the bottom of the different potential wells, indicating
the onset of a new dynamical instability.  In this case, however, a
Bogoliubov-like stability analysis is much more problematic because of
the non trivial temporal evolution of the condensate wave-packet.  A
possible, though approximate, approach is to write down an effective
Hamiltonian of the system in terms, for instance, of the collective
coordinates introduced in Sect.\ref{newton}.  Such Hamiltonian would
contain a limited number of degree of freedom, making the stability
analysis a much easier task.  This is the approach followed in
\cite{trombettoni01} to study the dynamics of an expanding condensate
in the discrete nonlinear Schr\"odinger equation framework (with
$m_\mu = m_\varepsilon$).  Within this approach one recovers, in a
unified framework, the critical values of the parameters for the
self-trapping conditions of a wavepacket of finite width initially at
rest, and the onset of the modulational instability of a Bloch wave
discussed in the previous sections. For instance, considering a 
gaussian wavepacket with initial width $\sigma_0$ and 
quasi-momentum $p_0$, the collective coordinates approach predicts the
onset of self-trapping at a  
critical value of the parameter $\Lambda=U_2 N_T/2K$ \cite{trombettoni01}. 
When $\cos{(p_0)}>0$, the critical value 
is $\Lambda_c \approx 2 \sqrt{\pi} \sigma_0 \cos{(p_0)}$; 
when $\cos{(p_0)}<0$, the critical value
is $\Lambda_c \approx
2 \sqrt{\pi} \mid \cos{(p_0)} \mid /\sigma_0$. We remark that 
when the width of the wavepacket is very large ($\sigma_0 \to \infty$), 
$\Lambda_c \to \infty$ if $\cos{(p_0)}>0$
(and the system is always dynamically stable), while 
$\Lambda_c \to 0$ if $\cos{(p_0)}<0$
(and the system is always dynamically unstable),
recovering the findings of Sect.IV.

The study of the dynamical instabilities of a condensate trapped in an
periodic potential is quite a rich problem, and deserves further
investigations. As we have mentioned, this is connected to the general
problem of the stability of a non stationary state, which includes for
instance also the propagation of sound waves in the non linear regime.
First experimental results on the self-trapping with weakly coupled
BECs are reported in \cite{morsch02,oberthaler}.

\section{CONCLUSIONS} 

The Gross-Pitaevskii dynamics of a Bose-Einstein condensate trapped in
a deep periodic potential can be studied in terms of a discrete,
nonlinear equation. This mapping allows a clear and intuitive picture
of the main dynamical properties of the system, which can be
calculated analytically. We have calculated the effective masses of
the system, connected to the Bloch energy and chemical potential
spectra. We have calculated the Bogoliubov dispersion relation, and
studied the sound velocity and the appearance of energetic and
dynamical instabilities. We have generalized these concepts to the
case of shallow optical lattice, which requires a numerical solution
and provides complementary insight in the understanding of the
problem.  Both in the tight binding limit and in the case of shallow
optical potential, we have investigated in detail the arising of
dynamical instabilities, which seem to be the main mechanism of
dephasing of the condensate in Bloch oscillation and dipole
oscillations experiments.

{\it Acknowledgements.}
We thank M. Kr\"amer, L. Pitaevskii and S. Stringari for interesting 
discussions. This work has been partially supported by the DOE.

{\it Note added in Proofs:} 
An equation similar to the DNL equation (\ref{DNLS_gen}) has been derived in 
M. \"Oster, M. Johansson, and A. Eriksson, Phys. Rev. E {\bf 67}, 
056606 (2003), to describe the dynamics of an electric field in an 
array of coupled optical waveguides embedded in a material with Kerr 
nonlinearities.

%\end{multicols}{2}

\end{document}